\documentclass[letters,useAMS,usenatbib]{mnras}
\usepackage{journals}
\usepackage{graphicx}
\usepackage{color}
\usepackage{times}
\usepackage{hyperref}
\usepackage{amsmath}
\usepackage{bm}
\usepackage{multirow}
\usepackage[warn]{textcomp}

\title[SNe~Ia distribution relative to spiral arms]{Constraining Type~Ia supernovae via their distances from spiral arms}
\author[A.~G.~Karapetyan]{Arpine~G.~Karapetyan\thanks{\selectfont{E-mail:
\href{mailto:a.karapetyan@yerphi.am}{a.karapetyan@yerphi.am}}}
\\
Center for Cosmology and Astrophysics, Alikhanian National Science Laboratory, 2 Alikhanian Brothers Str., 0036 Yerevan, Armenia}
\begin{document}

\date{Accepted 2022 October 5. Received 2022 September 29; in original form 2022 September 6}

\pagerange{\pageref{firstpage}--\pageref{lastpage}} \pubyear{2022}

\maketitle

\label{firstpage}

\begin{abstract}

  We present an analysis of the distribution of 77 supernovae (SNe) Ia relative to spiral arms of their
  Sab--Scd host galaxies, using our original measurements of the SN distances from the nearby arms,
  and study their light curve decline rates ($\Delta m_{15}$).
  For the galaxies with prominent spiral arms,
  we show that the $\Delta m_{15}$ values of SNe~Ia,
  which are located on the arms, are typically smaller (slower declining) than those of interarm SNe~Ia (faster declining).
  We demonstrate that the SN~Ia distances from the spiral arms and their galactocentric radii are correlated:
  before and after the average corotation radius, SNe~Ia are located near the inner and outer edges (shock fronts)
  of spiral arms, respectively.
  For the first time, we find a significant correlation between the $\Delta m_{15}$ values and SN distances from
  the shock fronts of the arms (progenitor birthplace), which is explained in the frameworks of
  sub-Chandrasekhar-mass white dwarf explosion models and density wave theory, where, respectively,
  the $\Delta m_{15}$ parameter and SN distance from the shock front are appropriate progenitor population age (lifetime) indicators.
\end{abstract}


\section{Introduction}
\label{intro}
\defcitealias{2016MNRAS.459.3130A}{A16}
\defcitealias{2020MNRAS.499.1424H}{H20}

Type Ia supernovae (SNe~Ia) are thought to be preceded by carbon-oxygen white dwarfs (WDs) in close binaries,
while the characteristics of the progenitors and the explosion channels are still up for debate \citep[e.g.][]{2018PhR...736....1L}.
It is now evident that SNe~Ia are not a homogeneous population of WD explosions,
instead they exhibit photometric and spectroscopic diversities \citep[e.g.][]{2017hsn..book..317T}.
One of the characteristic parameters of SNe~Ia is
the difference in $B$-band magnitudes between the max and 15 day,
or the so-called SN light curve (LC) decline rate $\Delta m_{15}$,
which is practically extinction-independent \citep[e.g.][]{1999AJ....118.1766P}.
There is a correlation between this parameter and SN~Ia luminosities at the maximum light:
SNe~Ia with larger $\Delta m_{15}$ values, or faster declining LCs, are fainter \citep{1993ApJ...413L.105P}.
The two most prevalent subclasses of peculiar SNe~Ia are 91bg-like events,
which are $\sim2$~mag subluminous at the $B$-band maximum than normal SNe~Ia and
have fast-declining LCs, and 91T-like SNe, which are $\sim0.6$~mag overluminous than normal ones
and have slow-declining LCs \citep[see][]{2017hsn..book..317T}.

Theoretically, in sub-Chandrasekhar-mass (sub-$M_{\rm Ch}<1.4 M_{\odot}$) explosion models, the luminosity of SN~Ia
is closely proportional to the exploding WD's mass
\citep[e.g.][]{2010ApJ...714L..52S,2017MNRAS.470..157B}:
WD in a double-degenerate system, which has mass lower than $M_{\rm Ch}$ may, under appropriate circumstances,
explodes as fainter SN~Ia with faster declining LCs
\citep{2017ApJ...851L..50S,2021ApJ...909L..18S}.
Note that, in comparison to WD around the $M_{\rm Ch}$ mass,
WD with a lower mass comes from a main-sequence progenitor star with a lower mass and
consequently with a longer lifetime (older progenitors).

From host galaxy studies, significant correlations are observed between SN~Ia LC decline rate and
the global ages of their hosts or local age at SN explosion sites
(e.g. \citealt{2011ApJ...740...92G,2014MNRAS.438.1391P,2016MNRAS.457.3470C,2020A&A...644A.176R};
\citealt[][hereafter \citetalias{2020MNRAS.499.1424H}]{2020MNRAS.499.1424H}).
On average, SNe~Ia with larger $\Delta m_{15}$ values are associated with older stellar environments.
On the other hand, important relationships between host stellar population and properties of SNe~Ia progenitors
can be found by looking at the distribution of SNe~Ia relative to spiral arms of galaxies
(e.g. \citealt[][hereafter \citetalias{2016MNRAS.459.3130A}]{2005AJ....129.1369P,2016MNRAS.459.3130A}).
It is worth noting that, according to the spiral density wave (DW) theory
\citep[][]{1964ApJ...140..646L,1969ApJ...158..123R},
star formation (SF) typically occurs at shock fronts near the edges of spiral arms.
From these regions, newly born SN progenitors move in the same direction as the disc rotation
with respect to the spiral pattern until they reach their explosion sites
(e.g. \citealt{2007AstL...33..715M}; \citetalias{2016MNRAS.459.3130A}).
The distance from the spiral arm/from progenitor birthplace is a potential indicator of SN~Ia progenitor lifetime
and thus can be used to constrain SN~Ia progenitors.
However, SN~Ia LC decline rates have never been examined in SN studies based on where they are
located on spiral arms or between, as well as $\Delta m_{15}$ as a function of the mentioned distance.
In this \emph{Letter}, we link the $\Delta m_{15}$ and SN~Ia distributions relative to spiral arms
of nearby host galaxies
and demonstrate, for the first time, how this could provide another interesting way
to study the properties of SN~Ia progenitors.

\section{Sample selection and reduction}
\label{samplered}

The database of this study consists of SNe~Ia from the sample of \citetalias{2020MNRAS.499.1424H},
after applying the restrictions described below.
Note that \citetalias{2020MNRAS.499.1424H} database is a compilation of
407 nearby SNe~Ia $(z \lesssim 0.036)$
with known spectroscopic subclasses and $B$-band LC decline rates $(\Delta m_{15})$.
In addition, the database contains information on the distances of SNe~Ia host galaxies,
morphologies, $ugriz$ magnitudes, and other parameters.
For the current study, we selected only normal, 91T-, and 91bg-like SN~Ia subclasses,
which include a sufficient number of events from a statistical perspective.

For hosts, we restricted to Sab--Scd morphologies since we are interested
in studying SNe~Ia in galaxies with well-developed arms,
where spiral DWs play an important role
\citep[e.g.][]{1987ApJ...314....3E,2016ApJ...827L...2P,2018MNRAS.481..566K}.
Following the approach of \citet{2014MNRAS.444.2428H},
we visually checked the levels of morphological disturbances of the host galaxies
using their images from different surveys.
The hosts with interacting and merging attributes were excluded from the sample
since we are interested in studying SNe~Ia in non-disturbed spiral galaxies.
In addition, to avoid projection and absorption effects in the discs due to high inclinations,
as well as to accurately investigate the immediate vicinity of SNe in terms of the existence
of host spiral arms, the sample is limited to galaxies with $i<60^\circ$.
Only 142 SNe~Ia in 137 host galaxies
met the applied restrictions.

In spiral galaxies, the vast majority of SNe~Ia belonging to normal, 91T-, and 91bg-like
subclasses are discovered in disc of hosts \citep[][]{2021MNRAS.505L..52H}.
Given this and using SN coordinates on the $g$-band images,\footnote{We used the FITS images from
the Sloan Digital Sky Survey (SDSS; \href{https://www.sdss.org/}{sdss.org}),
the SkyMapper (\href{https://skymapper.anu.edu.au/}{skymapper.anu.edu.au}),
and the Pan-STARRS (\href{https://outerspace.stsci.edu/display/PANSTARRS}{outerspace.stsci.edu/display/PANSTARRS})
surveys.}
for each SNe~Ia we visually inspected the area of a circular ring in a quadrant of host disc,
where the SN is discovered, in terms of presence of well-pronounced spiral arms.
This is important because we aimed to link SN progenitors to a population of stars born
due to the SF after passing and compressing gas clouds through the DW
(e.g. \citealt{2007AstL...33..715M}; \citetalias{2016MNRAS.459.3130A}).
SNe~Ia in the circumnuclear region or in the far outer disc
(SN galactocentric distance $< 0.1 \, R_{25}$ or $> R_{25}$)
were excluded from this
study,\footnote{$R_{25}$ is the $g$-band
$25^{\rm th}$ magnitude isophotal semi-major axis of the disc.}
as well as those visually identified within the radius swept up by host galactic bar.
As a result of this selection, we were finally left with 77 SNe~Ia in 74 host galaxies
(see Table~\ref{HOSTandSNIa}).

\begin{figure}
\begin{center}$
\begin{array}{@{\hspace{0mm}}c@{\hspace{2mm}}c@{\hspace{0mm}}}
  \includegraphics[width=0.46\hsize]{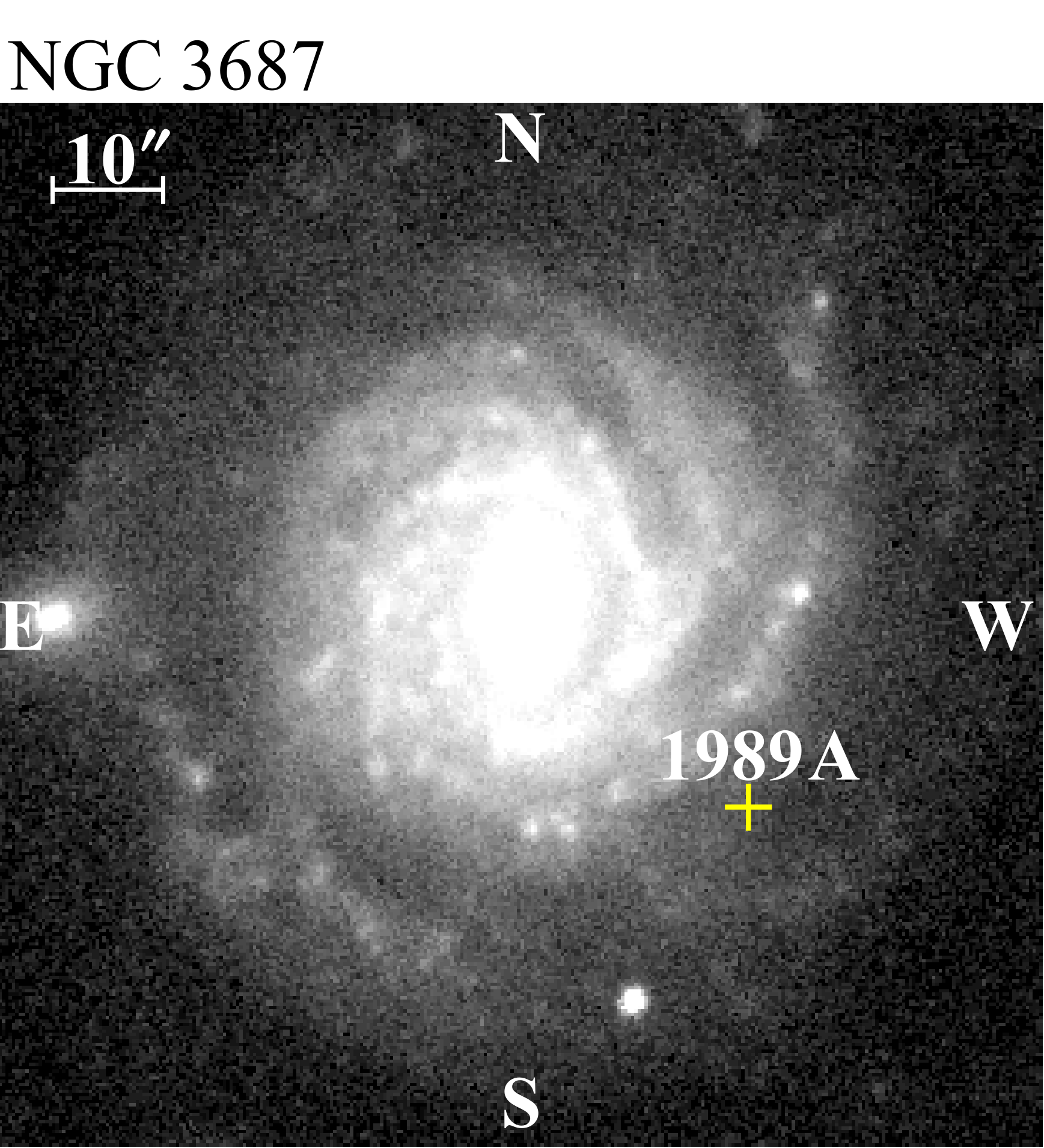} &
  \includegraphics[width=0.46\hsize]{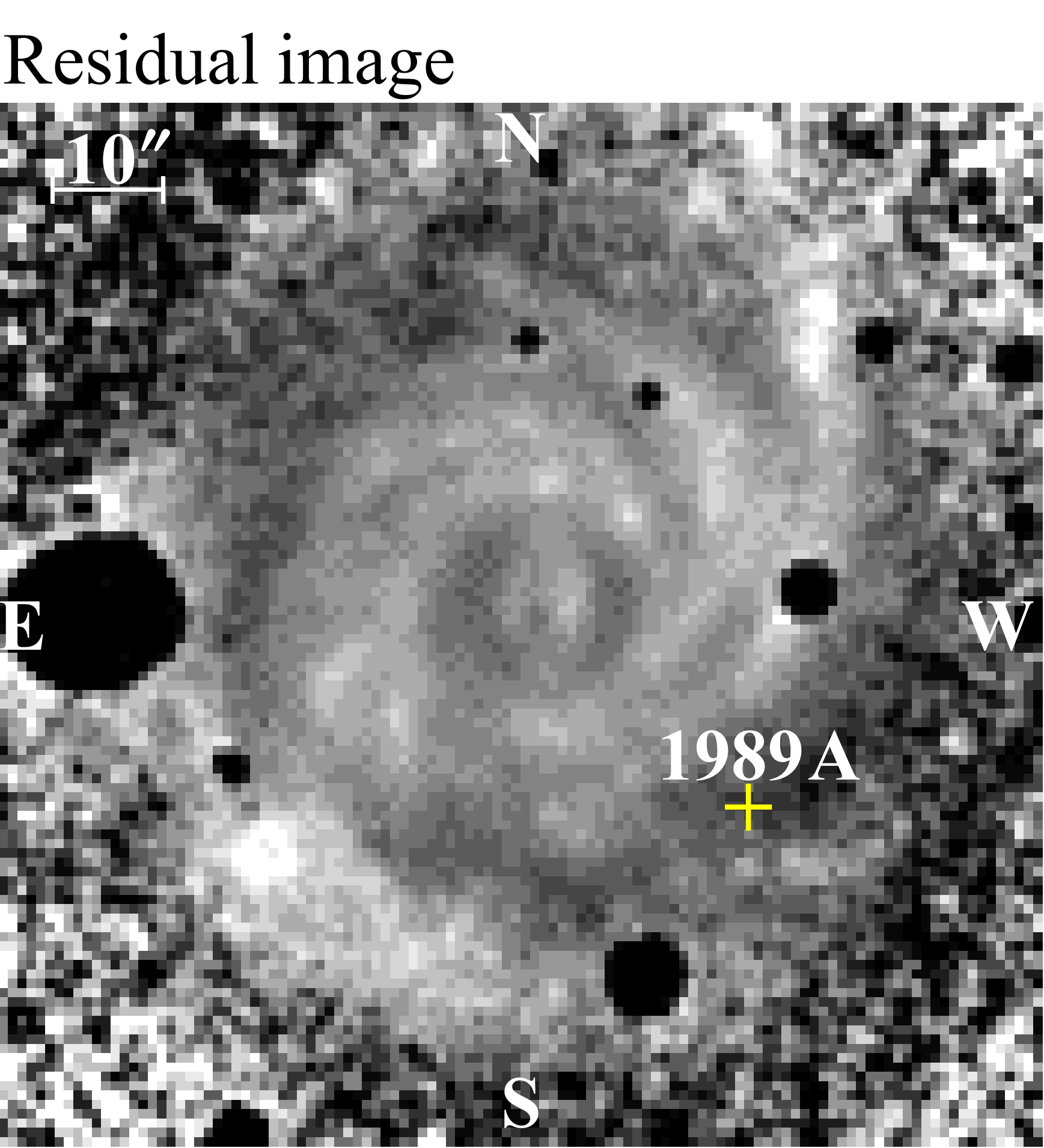} \\
  \includegraphics[width=0.46\hsize]{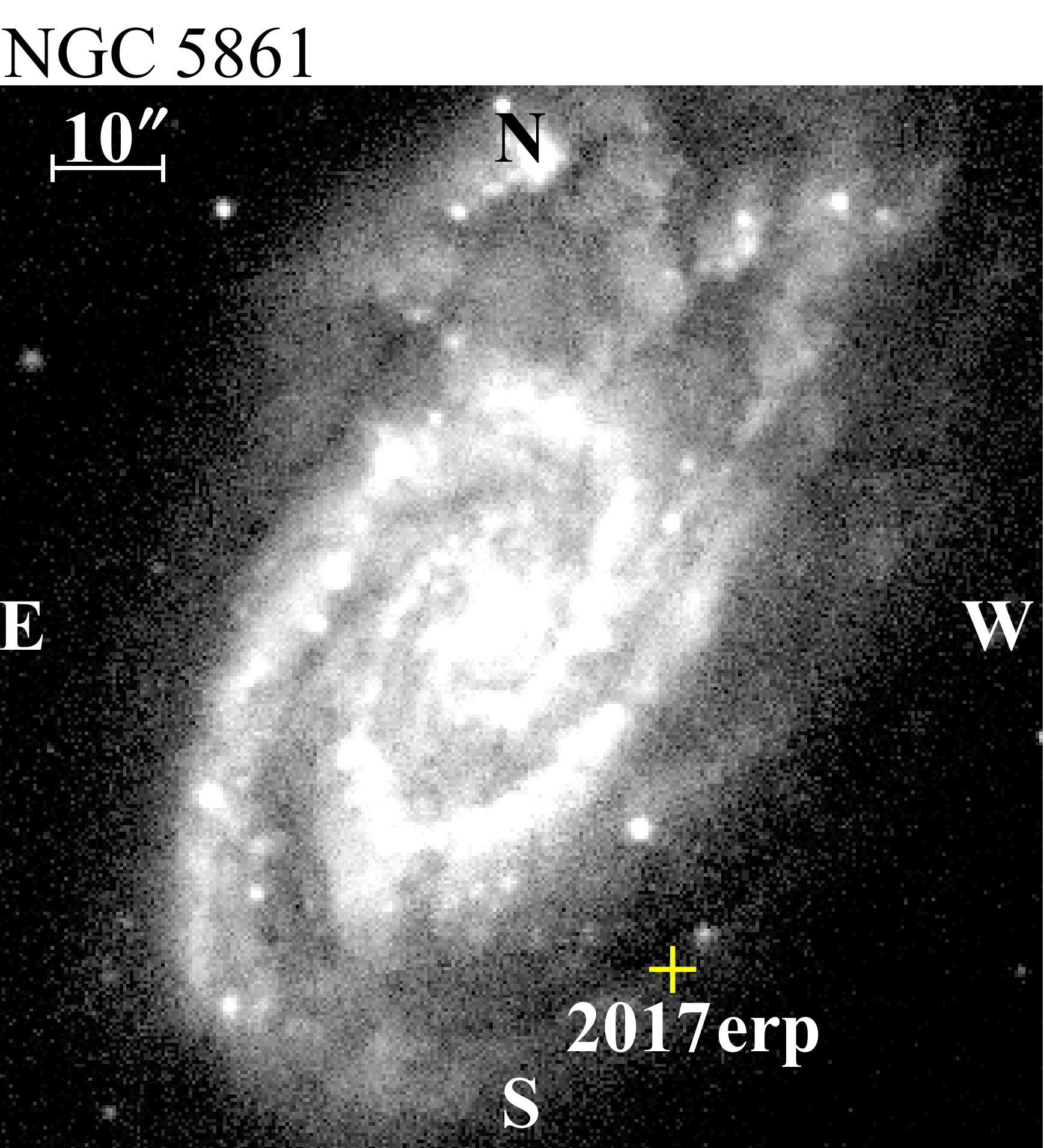} &
  \includegraphics[width=0.46\hsize]{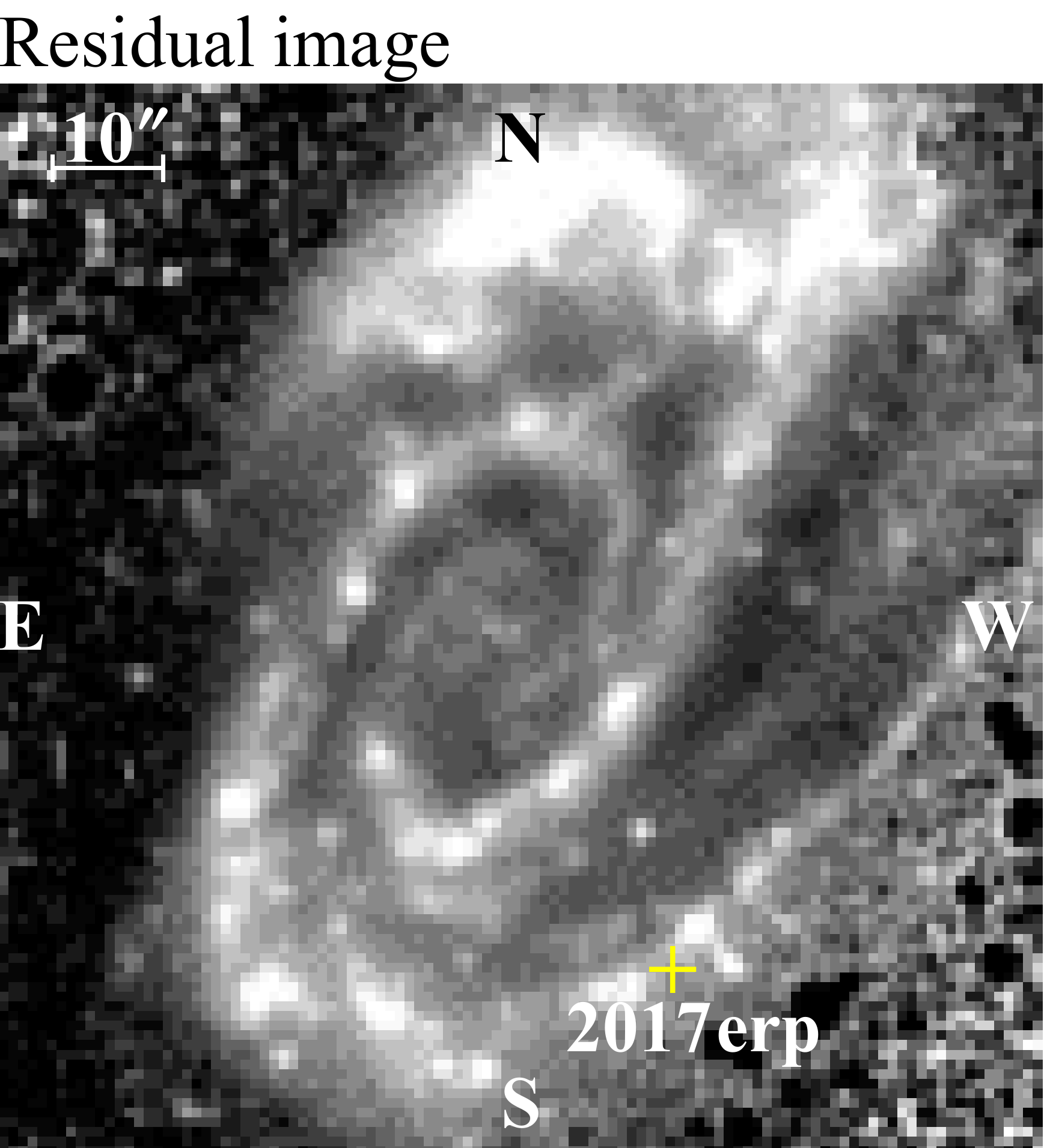}
\end{array}$
\end{center}
\caption{Upper panels: SDSS $g$-band image of \emph{interarm} SN~1989A host galaxy (left) and its residual image (right),
         after subtracting bulge and disc components.
         Bottom panels: Pan-STARRS $g$-band image of \emph{arm} SN~2017erp host (left) and its residual image (right).
         The locations of SNe are signed by crosses in all images (north is up and east to the left).
         Names of host galaxies are noted. In the residual images, bright projected stars are masked.}
\label{SubtractBulgeDisc}
\end{figure}

We determined the host spiral arm structures and the SNe positions with respect to
the spiral arms according to the approaches detailed in \citetalias{2016MNRAS.459.3130A}.
In short, we defined \emph{arm} and \emph{interarm} SNe~Ia
that are discovered inside the host arm edges or in the interarm region, respectively.
To accomplish this, we used the residual images of the host galaxies
after subtraction of the fitted $r^{1/4}$ bulge$+$exponential disc profiles
from the smoothed $g$-band fits images.
In the residual images, the values of the interarm pixels are negative,
since the fitted profiles use fluxes from both the arm and interarm regions.
Similar to \citetalias{2016MNRAS.459.3130A}, we fixed the edges of the spiral arms
when the flux values change the sign.
Fig.~\ref{SubtractBulgeDisc} shows examples of
original and bulge$+$disc subtracted images
of galaxies that host arm and interarm SNe~Ia.
Table~\ref{DW1DW2NDW1NDW2} lists the numbers of SNe~Ia in arm and interarm subsamples.

\begin{figure}
\begin{center}$
\begin{array}{@{\hspace{0mm}}c@{\hspace{0mm}}}
\includegraphics[width=\hsize]{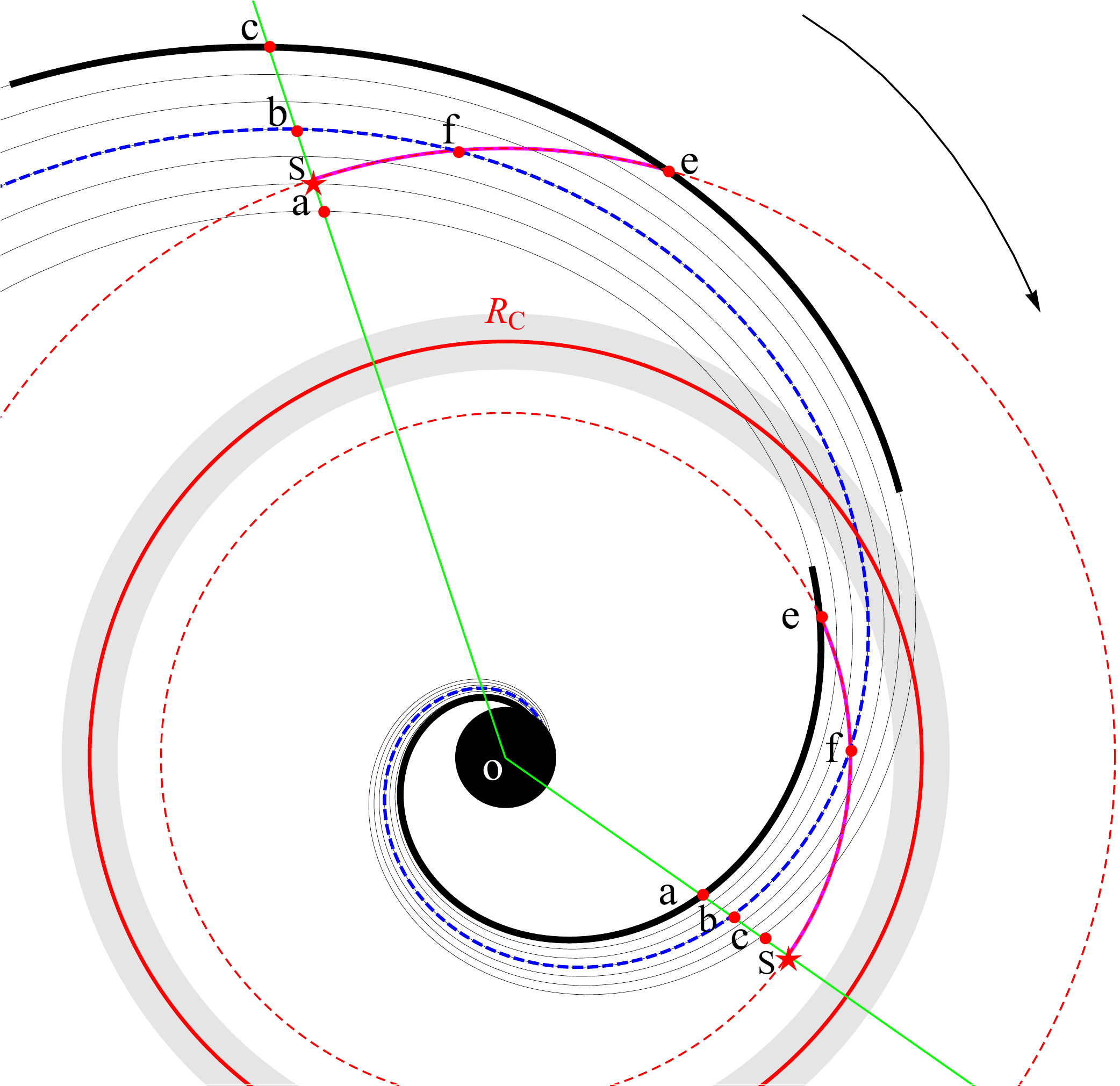}
\end{array}$
\end{center}
\caption{A scheme of grand-design galaxy with logarithmic arms (only one arm is shown),
         where additional SF is triggered by DWs.
         The arrow indicates the galaxy's rotational direction around nucleus (\texttt{O}).
         Thick black and blue dashed curves present shock fronts of spiral arm and arm's density (brightness) peak, respectively.
         A broad gray ring and a red solid circle indicate the corotation region and radius $R_{\rm C}$, respectively.
         Two orbits of SN progenitors are represented by red dashed circles.
         The purple arc depicts traveled distance of an SN progenitor from birthplace (\texttt{e})
         up to the explosion (\texttt{s}), through the arm peak (\texttt{f}) in the particular cases.
         The radial directions connecting SNe and nucleus are shown by green lines.
         The \texttt{ac}, \texttt{ba}, and \texttt{bc} lengths are full, inner, and outer
         (semi)widths of the arm, respectively.}
\label{SpArmSNfig}
\end{figure}

For each SNe~Ia in the subtracted images, we measured the distance $(d)$
from the $g$-band surface brightness peak
of the nearest spiral arm through the galactocentric direction
(\texttt{bs} distance in Fig.~\ref{SpArmSNfig}).
Only for two cases (SN1997cw and SN2002ck),
the interarm SN association with the nearest spiral arm is somewhat ambiguous.
Following \citetalias{2016MNRAS.459.3130A}, we normalized $d$ to the corresponding semiwidth
of the spiral arm, i.e. $\widetilde{d}=d/W_{\pm}$,
to compensate for the various linear sizes of the arm width.
The semiwidth is the distance from the spiral arm peak to the inner/outer edge
of the arm through the galactocentric direction.
The $W_{-}$ is the inner semiwidth (\texttt{ba} length in Fig.~\ref{SpArmSNfig}) with negative sign
when SN is located between the nearest spiral arm peak and the host galaxy nucleus,
and the $W_{+}$ is the outer semiwidth (\texttt{bc} length in Fig.~\ref{SpArmSNfig}) with positive sign
when the arm peak is between SN and the nucleus
(see \citetalias{2016MNRAS.459.3130A}, for details).

It is worth noting that, according to the DW theory \citep[e.g.][]{1964ApJ...140..646L,1969ApJ...158..123R},
SF activities usually take place at a shock front around the inner edges
of spiral arms inside the corotation radius $(R_{\rm C})$,
and around the outer edges of arms outside the corotation
(see Fig.~\ref{SpArmSNfig}).
For each SNe~Ia, we also measured the distance ($D$) from the shock front of
spiral arm through the galactocentric direction
(\texttt{as} and \texttt{cs} distances in Fig.~\ref{SpArmSNfig} inside and outside $R_{\rm C}$ radius, respectively).
We normalized $D$ to the width $(W)$ of the spiral arm (\texttt{ac} length in Fig.~\ref{SpArmSNfig}),
i.e. $\widetilde{D}=D/W$.

In addition, we estimated the deprojected galactocentric distances of SNe~Ia $(R_{\rm SN})$,
using well-known approach of correction for the host galactic disc inclination
\citep[see][for details]{2016MNRAS.456.2848H}.
This requires the offsets of SNe from the nucleus of host galaxies
$(\Delta\alpha$ and $\Delta\delta)$, the position angles and inclinations of the discs.
Eventually, for each SNe~Ia, the $R_{\rm SN}$ is normalized to the $g$-band $R_{25}$,
i.e. $\widetilde{r}_{\rm SN}=R_{\rm SN}/R_{25}$,
to compensate for the various linear sizes of hosts.
Note that the mentioned parameters are not listed in
\citetalias{2020MNRAS.499.1424H}, however they were compiled and/or estimated at
the time of that study and are now available in the online database of this \emph{Letter}.

Table~\ref{databasetab} contains new database of this paper on all
77 individual SNe~Ia (SN name, \emph{arm} and \emph{interarm} SN definition,
$\widetilde{d}$, $\widetilde{D}$, $\widetilde{r}_{\rm SN}$),
while \citetalias{2020MNRAS.499.1424H} contains data on
the spectroscopic subclasses and $B$-band $\Delta m_{15}$ of the events, as well as data on host galaxies.

\section{Results and discussion}
\label{RESults}

\subsection{SNe~Ia in \emph{arm} and \emph{interarm} regions of spiral galaxies}
\label{RESults1}

\begin{figure}
\begin{center}$
\begin{array}{@{\hspace{0mm}}c@{\hspace{0mm}}}
\includegraphics[width=\hsize]{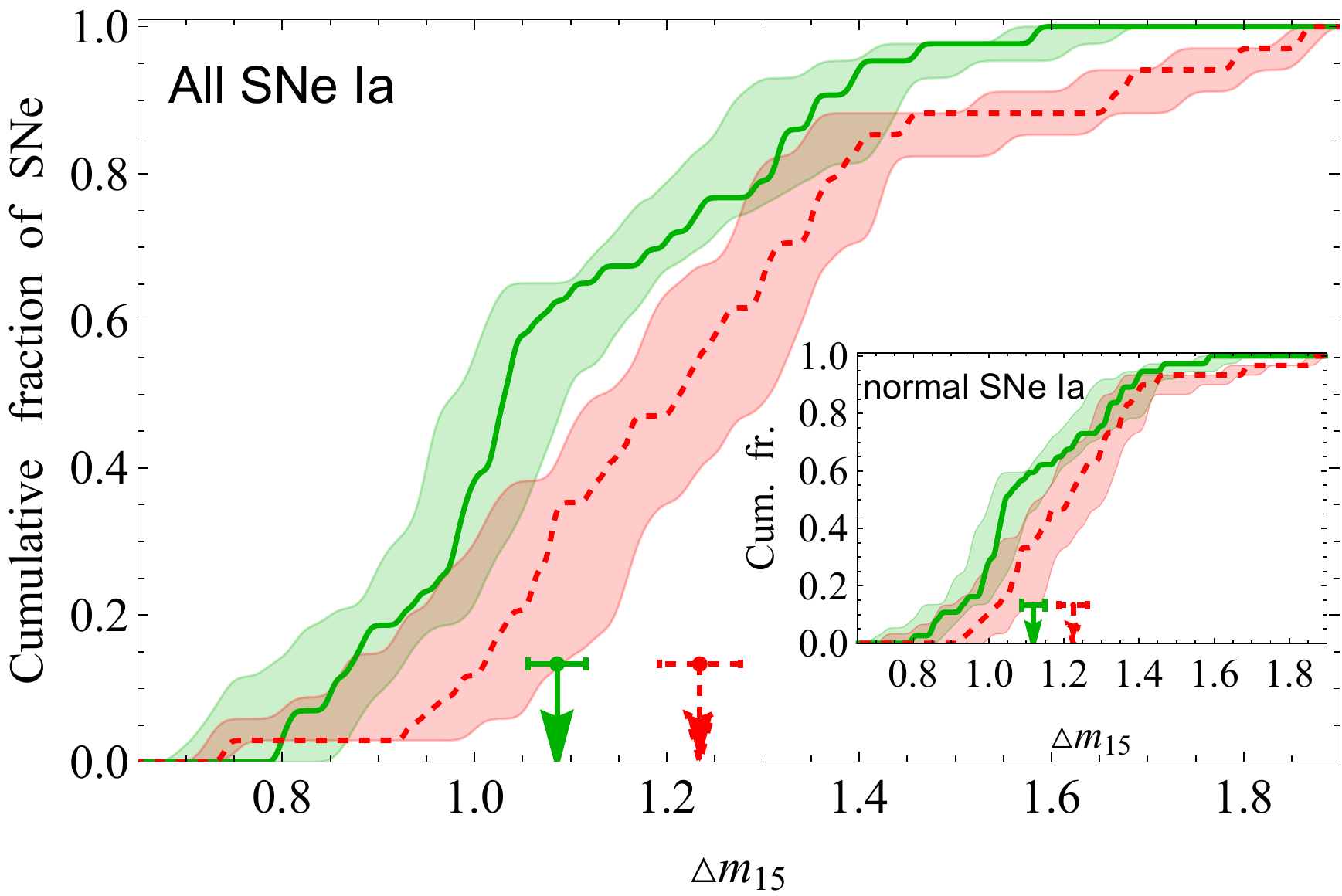}
\end{array}$
\end{center}
\caption{Cumulative $\Delta m_{15}$ distributions of all
         \emph{arm} (green solid) and \emph{interarm} SNe~Ia (red dashed).
         The associated spreads for each cumulative curve are shown by colored regions,
         taking into account the uncertainty in $\Delta m_{15}$ values.
         Arrows show the mean values (with their standard errors) of the distributions.
         The inset is the same but only for normal SNe~Ia.}
\label{Normdm15ArmInterarm}
\end{figure}

Fig.~\ref{Normdm15ArmInterarm} presents the cumulative $\Delta m_{15}$ distributions of
all sampled SNe~Ia in \emph{arm} and \emph{interarm} regions.
The inset in Fig.~\ref{Normdm15ArmInterarm} shows the same distributions,
but only for normal SNe~Ia.
To statistically compare the distributions, we use nonparametric methods
\citep[e.g.][]{Engmann+11}:
the two-sample Kolmogorov--Smirnov (KS) and Anderson--Darling (AD)
tests.\footnote{Due to the small number statistics,
to get a better estimate of the $P$-value of the KS and AD tests,
we employ Monte Carlo (MC) simulation with $10^5$ iteration,
as explained in detail in \citetalias{2020MNRAS.499.1424H}.
The threshold for the tests has traditionally been set at a 5 per cent significance level.}
The $P$-values of the tests in Table~\ref{S1S2PKSPAD}
indicate that the two $\Delta m_{15}$ distributions, being compared for all sampled SNe~Ia,
are significantly different.
For each SN~Ia subclass, we also try to perform the same comparison.
However, this can only be done for normal SNe~Ia, while the numbers are insufficient
for 91T- and 91bg-like events (see Table~\ref{DW1DW2NDW1NDW2}).
As for all SNe~Ia, the tests' results show that the $\Delta m_{15}$ distributions of
normal SNe~Ia in \emph{arm} and \emph{interarm} regions are inconsistent significantly
(with only barely AD test significance):
the $\Delta m_{15}$ values of \emph{arm} SNe~Ia are, on average, smaller (slower declining LCs)
in comparison with those of \emph{interarm} SNe~Ia (faster declining LCs).

\begin{table}
  \centering
  \begin{minipage}{84mm}
  \caption{Comparison of the LC decline rate distributions of \emph{arm} and \emph{interarm} SNe~Ia.}
  \tabcolsep 0.9pt
  \label{S1S2PKSPAD}
    \begin{tabular}{lccccccc}
    \hline
  \multicolumn{1}{l}{SN} & \multicolumn{1}{c}{$N_{\rm arm \, SN}$} & \multicolumn{1}{c}{$\langle \Delta m_{15} \rangle$} & \multicolumn{1}{c}{versus} & \multicolumn{1}{c}{$N_{\rm interarm \, SN}$} & \multicolumn{1}{c}{$\langle \Delta m_{15} \rangle$} & \multicolumn{1}{c}{$P_{\rm KS}^{\rm MC}$} & \multicolumn{1}{c}{$P_{\rm AD}^{\rm MC}$} \\
  \hline
   All & 43 & 1.09$\pm$0.03 & versus & 34 &  1.23$\pm$0.04 & \textbf{0.006} & \textbf{0.005}\\
   Normal & 37 & 1.12$\pm$0.03 & versus & 30 &  1.21$\pm$0.04 & \textbf{0.037} & 0.075\\
  \hline
  \end{tabular}
  \parbox{\hsize}{\emph{Notes.} The $\Delta m_{15}$ mean values and their standard errors for each sample are presented.
  MC simulation with $10^5$ iteration is used to provide $P_{\rm KS}^{\rm MC}$ and $P_{\rm AD}^{\rm MC}$
  probabilities for the KS and AD tests.
  Differences in the distributions that are statistically significant $(P\leq0.05)$ are marked in bold.}
  \end{minipage}
\end{table}

The results presented above can be interpreted
within the framework of DW theory \citep[e.g.][]{1964ApJ...140..646L,1969ApJ...158..123R}
and WD explosion models with a sub-$M_{\rm Ch}$
\citep[][]{2017ApJ...851L..50S,2021ApJ...909L..18S}.
According to the DW theory, stars (or SN~Ia progenitors) were born around shock fronts of spiral arms
(birthplace \texttt{e} in Fig.~\ref{SpArmSNfig})
and migrate in the same direction as the disc rotation relative to the spiral pattern
(traveled distance \texttt{eS}).
In comparison with arm SNe~Ia, interarm SNe~Ia should have, on average, longer lifetime of their progenitors
to travel from the birthplace through the entire arm and explode in interarm regions.
Therefore, it can be assumed that interarm SNe~Ia originates from an older progenitors than those in arms.
The arm/interarm separation thus provides an effective way to distinguish, on average, between
younger and older SN~Ia progenitors.

On the other hand, as mentioned in the Introduction, in sub-$M_{\rm Ch}$ explosion models
\citep[e.g.][]{2010ApJ...714L..52S,2017MNRAS.470..157B}
the $\Delta m_{15}$ of SN~Ia is correlated with the age of the progenitor system
\citep[larger $\Delta m_{15}$ values - older progenitors;][]{2017ApJ...851L..50S,2021ApJ...909L..18S}.
The described link, together with what is indicated in the paragraph above,
allows us to assume that interarm SNe~Ia come, on average, from older stellar population
with faster declining LCs in contrast to arm SNe~Ia.

\subsection{The distribution of SNe~Ia relative to spiral arms}
\label{RESults2}

To supplement and develop the results obtained in the previous section,
it is preferable to analyse continuous parameter distributions, such as the galactocentric radii of SNe
and their distances from the host spiral arm, and relate them with SN LC decline rates
rather than utilizing the \emph{arm} and \emph{interarm} discrete binning of SNe~Ia.

\begin{table}
  \centering
  \begin{minipage}{84mm}
  \caption{Results of Spearman's rank correlation tests for different continuous parameters of SNe~Ia.}
  \tabcolsep 9.2pt
  \label{spermanrank}
    \begin{tabular}{lccrr}
    \hline
  \multicolumn{1}{l}{SN} & \multicolumn{1}{l}{${N}_{\rm SN}$} & Par.~1 versus Par.~2 & \multicolumn{1}{c}{${r}_{\rm s}$} & \multicolumn{1}{c}{$P_{\rm s}^{\rm MC}$} \\
  \hline
   All & 77 & $\Delta m_{15} \,$ versus $\, \widetilde{r}_{\rm SN}$ & $0.032$ & 0.783\\
   Normal& 67 & $\Delta m_{15} \,$ versus $\, \widetilde{r}_{\rm SN}$ &$-0.021$ & 0.867 \\
   All & 77 & $\widetilde{d} \,$ versus $\, \widetilde{r}_{\rm SN}$ & 0.330 & \textbf{0.003} \\
   Normal & 67 & $\widetilde{d} \,$ versus $\, \widetilde{r}_{\rm SN}$ & 0.370 & \textbf{0.002} \\
  All & 77 & $\Delta m_{15} \,$ versus $\, \widetilde{D}$ & $0.288$ & \textbf{0.011} \\
  Normal& 67 & $\Delta m_{15} \,$ versus $\, \widetilde{D}$ & $0.280$ & \textbf{0.022} \\
   All & 77 & $\Delta m_{15} \,$ versus $\, |\widetilde{d}|$ & $0.183$ & 0.111 \\
   Normal& 67 & $\Delta m_{15} \,$ versus $\, |\widetilde{d}|$ & $0.077$ & 0.360 \\
  \hline
  \end{tabular}
  \parbox{\hsize}{\emph{Notes.} Spearman's coefficient $(-1 \leq {r}_{\rm s} \leq 1)$ is
                  a measure of rank correlation.
                  The test's null hypothesis is that the variables are independent,
                  whereas the alternate hypothesis is that they are not.
                  The permutations with $10^5$ MC iterations are used to generate the $P_{\rm s}^{\rm MC}$ values.
                  Statistically significant correlations are marked in bold $(P\leq0.05)$.}
  \end{minipage}
\end{table}

In this context, a negative radial gradient of stellar population age seen in spiral discs
\citep[e.g.][]{2015A&A...581A.103G}
prompts us to check the dependency between the $\Delta m_{15}$ and $\widetilde{r}_{\rm SN}$ of SNe~Ia.
This dependency has been studied extensively in the past, but no significant correlation has been found
\citep[e.g.][]{2005ApJ...634..210G, 2012ApJ...755..125G, 2017ApJ...848...56U,2021MNRAS.505L..52H}.
For different subsamples of our study,
the Spearman's rank test in Table~\ref{spermanrank} also shows not significant
trends between the mentioned parameters.
In our recent study \citep{2021MNRAS.505L..52H}, we explained this negative result by the observed fact
that in stacked spiral discs the azimuthally averaged stellar population age radially varies only from
around 10 down to 8.5~Gyr from the center to the periphery \citep[e.g.][]{2015A&A...581A.103G}.
While a significant correlation is observed between the LC decline rate and the global ages of hosts
(the ages range approximately from 1 to 10~Gyr,
\citealt{2011ApJ...740...92G,2014MNRAS.438.1391P,2016MNRAS.457.3470C};
\citetalias{2020MNRAS.499.1424H}).
The relatively narrower radial age range is most likely the reason why
the $\Delta m_{15}$ versus $\widetilde{r}_{\rm SN}$ correlation cannot be well-observed.

\begin{figure}
\begin{center}$
\begin{array}{@{\hspace{0mm}}c@{\hspace{0mm}}}
\includegraphics[width=\hsize]{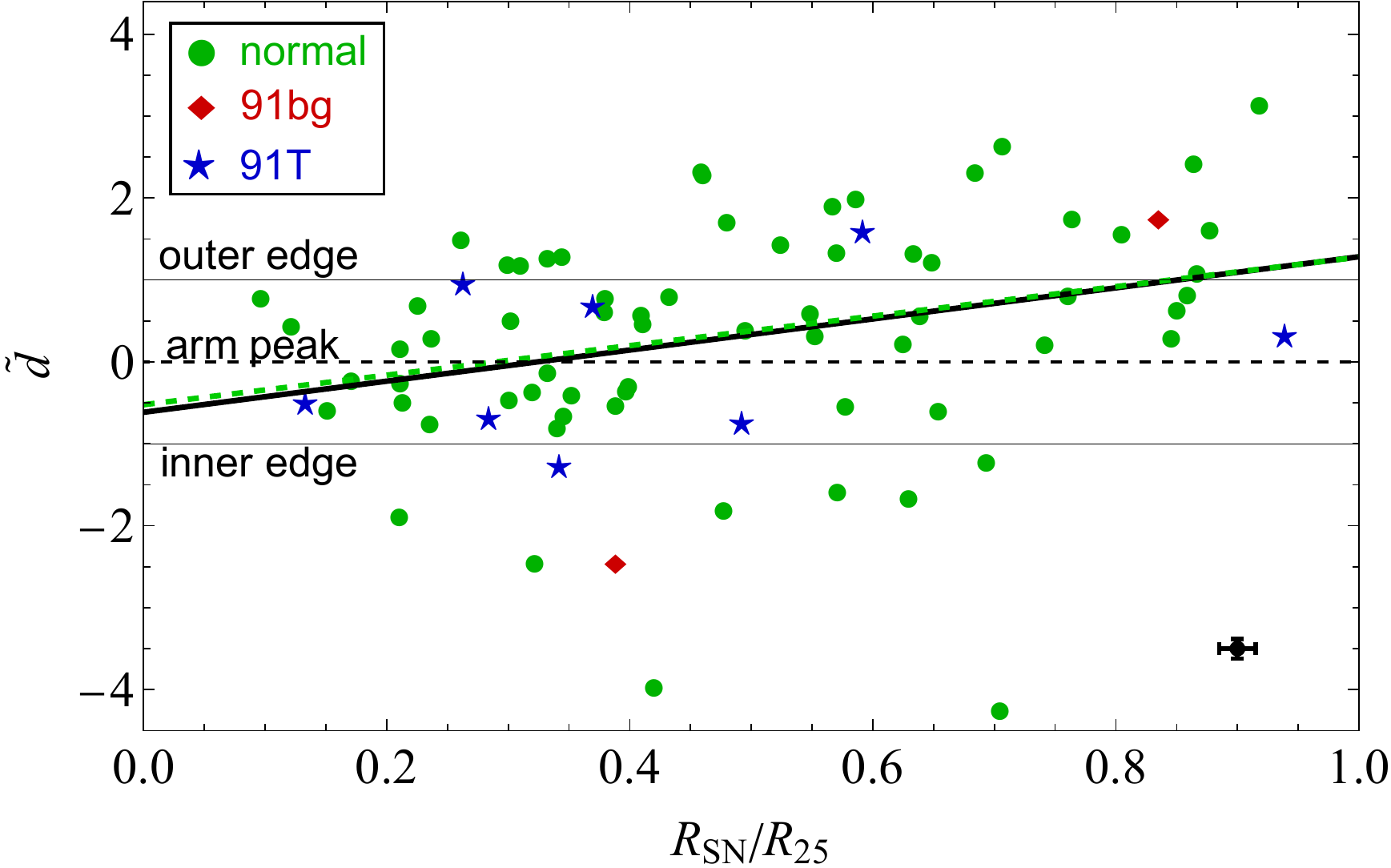}
\end{array}$
\end{center}
\caption{Distribution of the distances of SNe~Ia relative to the peaks of spiral arms versus the deprojected
         and normalized galactocentric distance.
         The inner and outer edges (solid lines), as well the peak of spiral arm (dashed line) are shown by parallel lines.
         The best fits for all and normal SN subclass are presented by the
         solid- and dashed-thick lines, respectively.
         The error bars in the bottom-right corner display the typical measurement errors.}
\label{rtildavsdtil}
\end{figure}

However, we can uncover an important relationship between host stellar population and properties of SNe~Ia
progenitors by looking at the distribution of SNe~Ia relative to spiral arms of galaxies
(e.g. \citealt{2005AJ....129.1369P,2007AstL...33..715M}; \citetalias{2016MNRAS.459.3130A}).
The relation between the normalized distances $\widetilde{d}$ of SNe~Ia from the arm peak and their
deprojected and normalized galactocentric distances $\widetilde{r}_{\rm SN}$ are shown in Fig~\ref{rtildavsdtil}.
There is a positive trend between the parameters, as shown by the fit line to the data.
The Spearman's rank correlation test in Table~\ref{spermanrank} indicates that
this trend is statistically significant for all and normal SNe~Ia samples.
In \citetalias{2016MNRAS.459.3130A}, the corresponding trend was not significant,
probably because of approximately 3.5 times smaller sample of SNe~Ia and
their different selection criteria for hosts and SN~Ia subclasses.

In Fig~\ref{rtildavsdtil}, the fit line to the distances of SNe~Ia relative to the arm peak intersects
with the arm roughly at a value of 0.35 in units of isophotal radii.
Since direct measurements of the corotation radii of host galaxies are not available for the current sample,
we examine the averaged value of $R_{\rm C}$ for seven host galaxies of SNe~Ia from our previous paper
\citet{2018MNRAS.481..566K}.
These galaxies' averaged morphological type, Sbc,
agrees well with that of the host galaxy sample used in the current study (Table~\ref{HOSTandSNIa}).
Moreover, the mean $R_{\rm C}\approx0.38\pm0.05$ for the mentioned sample from \citeauthor{2018MNRAS.481..566K}
is in good agreement with the intersection point in Fig~\ref{rtildavsdtil}.
Therefore, this intersection point 0.35 can be adopted as an average corotation radius for our hosts in units of isophotal radii.

The findings in Fig~\ref{rtildavsdtil} can be interpreted in the context of the DW theory where
the steady waves in grand-design galaxies have a strong influence on triggering SF processes
close to the shock fronts of spiral arms
\citep[e.g.][see also Fig.~\ref{SpArmSNfig}]{1964ApJ...140..646L,1969ApJ...158..123R}.
This is supported by the mentioned quantitative agreement for the average corotation radius of hosts and
the observational fact that the SNe~Ia explosion sites are mainly distributed around the inner and outer edges of
the arms (shock fronts) inside and outside the corotation radius, respectively.
Such locations of SNe~Ia may be due to a combination of the circular velocity of progenitors in the disc relative
to the pattern speed of the spiral arms (e.g. \citetalias{2016MNRAS.459.3130A}) and
the ages of SN~Ia progenitors \citep[e.g.][]{2014MNRAS.445.1898C}.
Long lived progenitors could travel farther by their circular orbits from the birthplaces around the shock fronts
to the explosion sites.
Inside the corotation radius this circular direction is from the inner to outer edges,
while outside the corotation the direction is from the outer to inner edges of arms
(e.g. \texttt{es} arcs in Fig~\ref{SpArmSNfig}).
Given that spiral galaxies are outnumbered by short-lived (prompt, i.e. 200-500 Myr) SN~Ia progenitors
\citep[e.g.][]{2009ApJ...707...74R,2014MNRAS.445.1898C},
we observe their distribution close to their birthplaces around the shock fronts.

From the above described DW scenario, we can assume that the traveled circular distance of
SN progenitor is an indicator of their age.
On the other hand, from the \texttt{esc} (\texttt{esa}) curvilinear triangle
outside (inside) the corotation in Fig~\ref{SpArmSNfig},
it can be understood that the \texttt{sc} (\texttt{sa}) distance is proportional to the
\texttt{es} distance.
Here, the \texttt{sc} (or \texttt{sa}) is the distance of SN~Ia from the shock front of
spiral arm through the galactocentric direction,
which we measured in Section~\ref{samplered} and normalized to the arm width (i.e. $\widetilde{D}$),
while the \texttt{es} is the traveled circular distance of SN progenitor.

\begin{figure}
\begin{center}$
\begin{array}{@{\hspace{0mm}}c@{\hspace{0mm}}}
\includegraphics[width=\hsize]{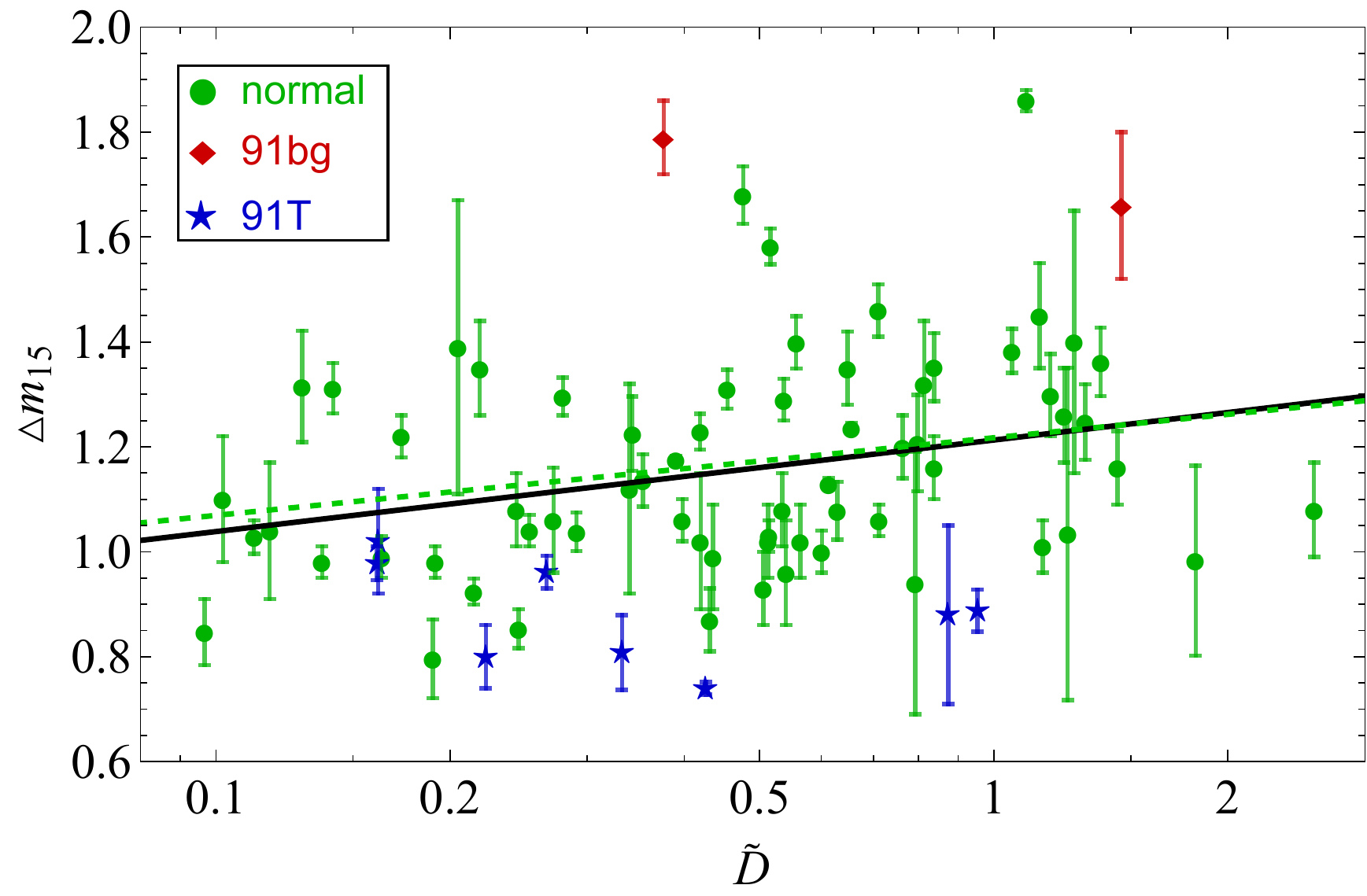}
\end{array}$
\end{center}
\caption{Distributions of $\Delta m_{15}$ values of SNe~Ia versus
         $\widetilde{D}$ distances from the shock fronts of host spiral arms.
         The best fits for all and normal SN subclass are presented by the
         solid- and dashed-thick lines, respectively.}
\label{Normdm15vsDtild}
\end{figure}

Considering that SNe~Ia LC decline rate can also be an age indicator for the progenitors in
sub-$M_{\rm Ch}$ explosion models,
in Table~\ref{spermanrank} and Fig~\ref{Normdm15vsDtild}
we study the correlations between $\Delta m_{15}$ values and
$\widetilde{D}$ distances from the shock fronts of host spiral arms.
The corresponding $P_{\rm s}$ values in Table~\ref{spermanrank} show that there
are significant correlations between these parameters.
The result enables us to draw the conclusion that, on average,
the progenitors of SNe~Ia with smaller $\Delta m_{15}$ values have shorter lifetimes and
thus traveled shorter distances from the shock fronts, i.e. birthplace,
in contrast to progenitors with larger $\Delta m_{15}$ values,
which have longer lifetimes and thus traveled farther away from the shock fronts.

The correlation tests in Table~\ref{spermanrank} show
the positive trend between the $\Delta m_{15}$ of SNe~Ia and their measured distances from the arm peak,
which might be assumed from the result of the $\Delta m_{15}$ differences between SN~Ia in arm/interam regions.
However, the $P_{\rm s}$ values of the test show that this trend is not statistically significant.
This insignificance likely caused by the blurs in $|\widetilde{d}|$ as a lifetime indicator in the
distribution of $\Delta m_{15}$ versus $|\widetilde{d}|$, because the SN distance from the arm peak does not represent
the progenitors' traveled distance during entire lifetime (till to SN explosion):
the spiral arm peak cannot be considered as a main birthplace of progenitors of SNe~Ia.

It is worth noting that when conducting all statistical tests of our study without
two SNe~Ia with ambiguous association with the nearest spiral arm (see Section~\ref{samplered}),
all the results of the \emph{Letter} and their significance remain unchanged.

\section{Conclusions}
\label{CONCL1}

In this \emph{Letter}, using a sample of Sab--Scd galaxies hosting 77 SNe~Ia
and our measurements of the SN distances from the nearby spiral arms,
we perform an analysis of the SNe distribution relative to host arms and
study their LC decline rates ($\Delta m_{15}$).
We demonstrate that the $\Delta m_{15}$ values of \emph{arm} SNe~Ia are typically
smaller (slower declining) than those of \emph{interarm} SNe~Ia (faster declining).
We show that the SN distances from the spiral arms and their galactocentric radii
are correlated: before and after the average corotation radius,
SNe~Ia are located near the inner and outer edges (shock fronts) of spiral arms.
For the first time, we find a correlation between $\Delta m_{15}$ values
and the SN distances from the shock fronts of the arms.
The results can be interpreted within the frameworks of DW theory,
where SN progenitors were born around shock fronts of spiral arms and
migrate crossing the spiral pattern to the explosion sites,
and WD explosion models with sub-$M_{\rm Ch}$,
where SN LC decline rate is an indicator of progenitor age.
On average, the progenitors of SNe~Ia with smaller $\Delta m_{15}$ values
have shorter lifetimes and thus traveled shorter distances from the shock fronts,
i.e. birthplace, in contrast to progenitors with larger $\Delta m_{15}$ values,
which have longer lifetimes and thus traveled farther away from the shock fronts.

As our study used a small sample size,
we encourage new statistically more powerful studies with larger and
more robust datasets of SNe~Ia and their hosts
(e.g. integral field observations, with available corotation radii)
to better constrain the nature of SN~Ia progenitor.

\section*{Acknowledgements}

We appreciate the anonymous referee for
comments towards improving our \emph{Letter}.
I am grateful for the help of my PhD supervisor, Dr.~Artur Hakobyan.
The work was supported by the Science Committee of RA,
in the frames of the research project \textnumero~21T--1C236.


\section*{Data Availability}
The online version of this \emph{Letter} contains the data
underlying this study (see Table~\ref{databasetab}, for instructions).


\bibliography{SpArmSNIabib}

\appendix
\section{Online material}

Table~\ref{HOSTandSNIa} shows morphological distribution of the sampled SNe~Ia
host galaxies, split between different SN subclasses.
The numbers of \emph{arm} and \emph{interarm} SNe~Ia within Sab--Scd galaxies
are shown in Table~\ref{DW1DW2NDW1NDW2}.

\begin{table}
  \centering
  \begin{minipage}{84mm}
  \caption{Morphological distribution of the sampled SNe~Ia host galaxies,
           split between different SN subclasses.}
  \tabcolsep 11pt
  \label{HOSTandSNIa}
    \begin{tabular}{lcccccc}
    \hline
  \multicolumn{1}{l}{SN} & \multicolumn{1}{c}{Sab} & \multicolumn{1}{c}{Sb} & \multicolumn{1}{c}{Sbc} & \multicolumn{1}{c}{Sc} & \multicolumn{1}{c}{Scd} & \multicolumn{1}{c}{All}\\
    \hline
    Normal & 6 & 18 & 25 & 17 & 1 & 67 \\
    91T & 2& 0 & 4 & 2 & 0 & 8 \\
    91bg & 2 & 0 & 0 & 0 & 0 & 2 \\
    ALL & 10 & 18 & 29 & 19 & 1 & 77 \\
  \hline
  \end{tabular}
  \end{minipage}
\end{table}
\begin{table}
  \centering
  \begin{minipage}{84mm}
  \caption{Numbers of \emph{arm} and \emph{interarm} SNe~Ia within Sab--Scd galaxies,
           split between different SN subclasses.}
  \tabcolsep 29.3pt
  \label{DW1DW2NDW1NDW2}
    \begin{tabular}{lcc}
    \hline
  \multicolumn{1}{l}{SN} & \multicolumn{1}{c}{\emph{arm}} & \multicolumn{1}{c}{\emph{interarm}}\\
  \hline
   Normal & 37 & 30 \\
   91T & 6 & 2 \\
   91bg & 0 & 2 \\
   All & 43 & 34 \\
  \hline
  \end{tabular}
  \end{minipage}
\end{table}

Table~\ref{databasetab} shows the first ten rows of the data used in our \emph{Letter}.
A CSV file containing the whole table is available online.

\begin{table}
\centering
\begin{minipage}{84mm}
  \caption{The database of 77 SNe~Ia of the study.
           The first ten entries are presented,
           while the entire table are available online.}
  \tabcolsep 8.5pt
  \label{databasetab}
  \begin{tabular}{lccccc}
    \hline
    \multicolumn{1}{c}{SN} & \multicolumn{1}{c}{Arm/interarm definition} & \multicolumn{1}{c}{$\widetilde{d}$} & \multicolumn{1}{c}{$\widetilde{D}$} & \multicolumn{1}{c}{$\widetilde{r}_{\rm SN}$} \\
    \hline
     1989A & interarm & 1.442 & 0.272 & 0.524 \\
     1989B & arm & 0.220 & 0.420 & 0.171 \\
     1990N & arm & 0.830 & 0.117 & 0.859 \\
     1990O & interarm & 1.753 & 0.541 & 0.764 \\
     1995al & arm & 0.516 & 0.601 & 0.302 \\
     1995E & arm & 0.478 & 0.398 & 0.411 \\
     1996ai & arm & 0.584 & 0.163 & 0.151 \\
     1996bo & arm & 0.445 & 0.537 & 0.122 \\
     1996Z & interarm & 1.226 & 0.173 & 0.649 \\
     1997bp & interarm & 2.326 & 0.837 & 0.459 \\
    \hline
  \end{tabular}
\end{minipage}
\end{table}

\label{lastpage}

\end{document}